\documentstyle[eqsecnum,preprint,prd,aps,psfig]{revtex}
\tighten
\let\jnfont=\rm
\def\MPLA#1,{{\jnfont Mod.\ Phys.\ Lett.\ A }{\bf #1},}
\def\NPB#1,{{\jnfont Nucl.\ Phys.\ }{\bf B#1},}
\def\PLB#1,{{\jnfont Phys.\ Lett.\ B }{\bf #1},}
\def\PRD#1,{{\jnfont Phys.\ Rev.\ D }{\bf #1},}
\def\PRL#1,{{\jnfont Phys.\ Rev.\ Lett.\ }{\bf #1},}
\def\PRep#1,{{\jnfont Phys.\ Rep.\ }{\bf #1},}
\def\ZPC#1,{{\jnfont Z.~Phys.\ C }{\bf #1},}
\begin{document}

\preprint{\parbox{2.0in}{\noindent TU-569\\ RCNS-99-04\\ AMES-HET-99-07\\
 July 1999\\}}

\title{\ \\[10mm] $R$-parity violation and top quark polarization 
at the Fermilab Tevatron collider}

\author{\ \\[2mm] Ken-ichi Hikasa and Jin Min Yang}

\address{\ \\[2mm] 
{\it Department of Physics, Tohoku University,
     Aoba-ku, Sendai 980-8578, Japan}}
               
\author{\ \\[2mm] Bing-Lin Young}

\address{\ \\[2mm] 
{\it Department of Physics and Astronomy, Iowa State University,
     Ames, Iowa 50011, USA}}
\maketitle

\begin{abstract}
The lepton or baryon number violating top quark interactions in the 
supersymmetric standard model with $R$ parity violation contribute to 
the process $d\bar d \to t\bar t$ at the tree level 
via the $t$- or $u$-channel sfermion exchange.  
Since these interactions are chiral, they induce polarization to the 
top quark in the $t\bar t$ events at hadron colliders.  We show 
in this article that the polarization can be a useful observable for probing 
these interactions at the upgraded Fermilab Tevatron collider,  because 
the polarization is expected to be very small in the standard model.  
\end{abstract}


\section{Introduction}
\label{sec:Introduction}

The top quark has an exceptionally large mass of the order of the electroweak 
symmetry breaking scale and is naturally expected to have a close 
connection to new physics~\cite{NP}.  
The Run~1 of the Fermilab Tevatron collider has successfully ended with the 
discovery of the top quark, but has yielded only a relatively small number 
of top quark events, which leaves plenty of room for new physics to be 
uncovered with the Tevatron upgrade~\cite{tev2000} in the near future.  
Due to higher statistics, the $t\bar t$ sample at the upgraded Tevatron 
is expected to provide a sensitive probe of new physics~\cite{tt1,tt2}. 

A most popular model of new physics is the Minimal Supersymmetric Model 
(MSSM){}.  In this model, the discrete multiplicative symmetry of 
$R$-parity, defined by $R=(-1)^{2S+3B+L}$ with spin $S$, baryon number $B$ 
and lepton number $L$, is often imposed on the Lagrangian to maintain the 
separate conservation of $B$ and $L$.  
This conservation requirement is, however, not dictated by any fundamental 
principle such as gauge invariance or renormalizability. 
The finiteness of the neutrino mass as suggested by the Super-Kamiokande 
and several other neutrino experiments~\cite{SK} 
also implies that lepton numbers may be violated.%  
\footnote{In fact, the $L$-violating interactions of the $R$-violating 
MSSM can give rise to neutrino masses at the one-loop level~\cite{rvmssm}.} 

The most general superpotential of the MSSM consistent with the 
$\rm SU(3)\times SU(2)\times U(1)$ symmetry and supersymmetry 
contains $R$-violating interactions which are given by~\cite{rvmssm}
\begin{eqnarray}\label{WR}
{\cal W}_{\not \! R}
=\frac{1}{2}\lambda_{ijk} L_i L_j E_k^c
+\lambda_{ijk}' \delta^{\alpha\beta} L_i Q_{j\alpha} D_{k\beta}^c
+\frac{1}{2} \lambda_{ijk}'' \epsilon^{\alpha\beta\gamma}
U_{i\alpha}^c D_{j\beta}^c D_{k\gamma}^c +\mu_i L_i H_2 .
\end{eqnarray}
Here $L_i$ ($Q_i$) and $E_i$ ($U_i$, $D_i$) are the left-handed 
lepton (quark) doublet and right-handed lepton (quark) singlet chiral 
superfields, and $c$ denotes charge conjugation.  
$H_{1,2}$ are the Higgs chiral superfields.  
The indices $i$, $j$, $k$ denote generations and 
$\alpha$, $\beta$ and $\gamma$ are the color indices.  
The $\lambda$ and $\lambda'$ are the coupling constants of 
the $L$-violating interactions and $\lambda''$ those of the 
$B$-violating interactions.  The non-observation (so far) of the proton decay 
imposes very strong constraints on the product of the $L$-violating 
and $B$-violating couplings~\cite{proton}.  It is thus conventionally assumed 
in phenomenological studies that only one type of these interactions 
(either $L$- or $B$-violating) exists.  
Constraints on these $R$-parity violating couplings have been obtained 
from various analyses~\cite{uni,nue,apv,Zdecay,Agashe,Kdecay,other} and 
some of their phenomenological implications at lepton~\cite{lepton} 
and hadron~\cite{hadron} colliders have been investigated.  
It is notable that the bounds on the couplings involving top quark are 
generally quite weak (see Ref.~\cite{review} for a review).  
In the future precision top quark experiments at the upgraded Tevatron, 
these couplings may either manifest themselves or subjected to
further, stronger constraints. 

In proton-antiproton collisions, the $L$-violating top quark interaction 
$\lambda'$ can induce tree-level processes such as 
$d\bar d\rightarrow t_L \bar t_L$ via slepton exchange in the $t$ channel.
(Here the subscript $L$ stands for chirality, not helicity.)
Similarly, the $B$-violating $\lambda''$ coupling gives rise to 
the process $d\bar d\rightarrow t_R \bar t_R$ via squark exchange.  
Since the couplings are chiral, 
these new production mechanisms produce an asymmetry between the 
left- and right-handed polarized top quarks. 
This polarization can be a sensitive probe for these couplings 
due to the fact that both the SM and $R$-conserving 
MSSM contributions to the polarization are small%
\footnote{For $\tan\beta<1$ and 
light gluino, which is disfavored by the existing experimental data, 
the polarization induced by supersymmetric weak and strong interactions
could be significant, however~\cite{pari}.}. 

In Ref.~\cite{tt}, the contribution of these processes 
to the {\it total\/} $t\bar t$ cross section was studied to constrain the 
relevant $R$-violating couplings. 
The effect is found to be at most of the level of a few percent.   
Because of the uncertainties in the SM predictions of the $t\bar{t}$ 
production cross section and possible contributions 
from the $R$-{\it conserving\/} MSSM interactions~\cite{loop}, 
the total $t\bar t$ cross section turns out not to be a suitable 
observable for the R-parity violating interactions. 
In this paper, we investigate the possibility of probing the $R$-violating 
in several observables including the top quark polarization in 
$t\bar t$ production at the upgraded Tevatron collider.  
The top polarization can serve as a sensitive probe of the R-parity violating 
interactions because of the negligibly small SM or MSSM contribution.

\section{Probing $R$-violation via top polarization}
\label{sec2}
In terms of the four-component Dirac notation, 
the Lagrangian of the $\lambda'$ and $\lambda''$ 
interactions is given by 
(notice that $\lambda''_{ijk}=-\lambda''_{ikj}$)
\begin{eqnarray}\label{lag1}
{\cal L}_{\lambda'}&=&-\lambda'_{ijk}
\left [\tilde \nu^i_L\bar d^k_{R\alpha} d^{j\alpha}_L
       +\tilde d^{j\alpha}_L\bar d^k_{R\alpha}\nu^i_L
    +(\tilde d^{k\alpha}_R)^*(\bar \nu^i_L)^c d^{j\alpha}_L\right.\nonumber\\
& & \qquad\quad \left. -\tilde e^i_L\bar d^{k}_{R\alpha} u^{j\alpha}_L
       -\tilde u^{j\alpha}_L\bar d^k_{R\alpha} e^i_L
       -(\tilde d^{k\alpha}_R)^*(\bar e^i_L)^c u^{j\alpha}_L\right ] 
       + {\rm h.c.},\\
\label{lag2}
{\cal L}_{\lambda''}&=&-\lambda''_{ijk}
\epsilon_{\alpha\beta\gamma}
\left [(\tilde d^{k\gamma}_R)^* \bar u^{i\alpha}_R (d^{j\beta}_R)^c
 +\frac12 (\tilde u^{i\alpha}_R)^* \bar d^{j\beta}_R (d^{k\gamma}_R)^c\right ]
       + {\rm h.c.}
\end{eqnarray}
The relevant Feynman diagrams for the process $d\bar d \to t\bar t$ 
via the $L$- and $B$-violating couplings are shown in Fig.~1.  
It is evident that these amplitudes are proportional to 
$|\lambda'_{i31}|^2$ and $|\lambda''_{31i}|^2$, respectively.  
The form of these amplitudes are discussed in some detail in the 
Appendix.  Here we show the cross section at $|\lambda|^2$ (the 
interference term with the SM $O(\alpha_s^2)$ amplitude), 
$d\hat\sigma^{\rm new}_{\lambda_t \lambda_{\bar t}}/d\cos\theta^*$ 
for each $t$, $\bar t$ helicity state $\lambda_{t,\bar t}=\pm$.  
For the $L$-violating couplings, we find 
\begin{mathletters}
\begin{eqnarray}
{d\hat\sigma^{\rm new}_{+-}\over d\cos\theta^*} 
&=& -{\alpha_s |\lambda'_{i31}|^2 \beta\over 72 \hat s}\,
{(1-\beta) (1+\cos\theta^*)^2 
\over 1+2(m_{\tilde e^i}^2-m_t^2)/\hat s - \cos\theta^*}\;,\\[3\jot]
{d\hat\sigma^{\rm new}_{++}\over d\cos\theta^*} =
{d\hat\sigma^{\rm new}_{--}\over d\cos\theta^*} 
&=& -{\alpha_s |\lambda'_{i31}|^2 \beta m_t^2\over 18 \hat s^2}\,
{\sin^2\theta^*
\over 1+2(m_{\tilde e^i}^2-m_t^2)/\hat s - \cos\theta^*}\;,\\[3\jot]
{d\hat\sigma^{\rm new}_{-+}\over d\cos\theta^*} 
&=& -{\alpha_s |\lambda'_{i31}|^2 \beta\over 72 \hat s}\,
{(1+\beta) (1-\cos\theta^*)^2 
\over 1+2(m_{\tilde e^i}^2-m_t^2)/\hat s - \cos\theta^*}\;,
\end{eqnarray}
\end{mathletters}
where $\hat s$ is the $t\bar t$ c.m.\ energy squared, $\theta^*$ 
the scattering angle in the $t\bar t$ c.m.\ frame, and 
$\beta = (1-4m_t^2/\hat s)^{1/2}$.  A sum over the flavor index 
$i$ is implied.  

Similar processes $s\bar s\to t\bar t$ and $b\bar b\to t\bar t$ 
occurs through the couplings $\lambda'_{i32}$ and $\lambda'_{i33}$, 
respectively.  The amplitudes for these processes are given by 
the same formulas with obvious substitution of the coupling.  
We will neglect these processes as the parton distributions of 
$s$ and $b$ quarks are small in a proton.

The corresponding expressions for the SM (QCD) lowest-order amplitude 
is 
\begin{mathletters}
\begin{eqnarray}
{d\hat\sigma^0_{+-}\over d\cos\theta^*} =
{d\hat\sigma^0_{-+}\over d\cos\theta^*} 
&=& {\pi\alpha_s^2 \beta\over 18 \hat s}\,
(1+\cos^2\theta^*)\;,\\[3\jot]
{d\hat\sigma^0_{++}\over d\cos\theta^*} =
{d\hat\sigma^0_{--}\over d\cos\theta^*} 
&=& {2\pi\alpha_s^2 \beta m_t^2\over 9 \hat s^2}\,
\sin^2\theta^*\;.
\end{eqnarray}
\end{mathletters}
The sum of these two contributions give the cross section 
$d\hat\sigma_{\lambda_t\lambda_{\bar t}}/d\cos\theta^*$ up to 
order $|\lambda|^2$.  

When only the polarization of the top quark is observed, we can sum 
over the $\bar t$ helicity
\begin{equation}
{d\hat\sigma_{\lambda_t} \over d\cos\theta^*} 
= {d\hat\sigma_{\lambda_t +} \over d\cos\theta^*} 
+ {d\hat\sigma_{\lambda_t -} \over d\cos\theta^*} \;.
\end{equation}
Integrating over the scattering angle, we obtain
\begin{eqnarray}
\hat{\sigma}_+^0 = \hat\sigma_-^0
&=& {4\pi\alpha_s^2\beta\over 27\hat s}\left(1+{2m_t^2\over \hat s}\right)
\;,\\
\label{LV1}
\hat\sigma_+^{\rm new} + \hat\sigma_-^{\rm new}
&=& \frac{\alpha_s |\lambda'_{i31}|^2 \beta}{9\hat s}
    \left\{ -\frac12 + {m_{\tilde e^i}^2-m_t^2\over \hat s}\right.\nonumber\\
&& \left. -{1\over\beta}\left[{m_t^2\over\hat s} 
  + {\bigl(m_{\tilde e^i}^2-m_t^2\bigr)^2 \over \hat s^2} \right]\,
     \log{(1+\beta)^2+4m_{\tilde e^i}^2/\hat s \over 
     (1-\beta)^2+4m_{\tilde e^i}^2/\hat s } \right\}\\
\label{LV2}
\hat\sigma_+^{\rm new} - \hat\sigma_-^{\rm new}
&=& \frac{\alpha_s |\lambda'_{i31}|^2}{9\hat s}
    \left\{ \frac12 - {m_{\tilde e^i}^2-m_t^2\over \hat s} \right.\nonumber\\
&& \left. -{1\over\beta}\left[{m_t^2\over\hat s}
  - {\bigl(m_{\tilde e^i}^2-m_t^2\bigr)^2 \over \hat s^2} \right]\,
     \log{(1+\beta)^2+4m_{\tilde e^i}^2/\hat s \over 
     (1-\beta)^2+4m_{\tilde e^i}^2/\hat s } \right\}
\end{eqnarray}

We define several observables sensitive to the $R$-violating interactions.  
The (total) top polarization is given by
\begin{equation}
P_t = {\hat\sigma_+ - \hat\sigma_- \over \hat\sigma_+ + \hat\sigma_-} \;.
\end{equation}
The hemisphere top polarizations are obtained 
by restricting the top scattering angle in the $t\bar t$ c.m.\ frame 
to the forward or backward hemisphere.  
The forward top polarization is defined as
\begin{equation}
P_t^F = {\hat\sigma_+(\cos\theta^*>0) - \hat\sigma_-(\cos\theta^*>0) 
\over \hat\sigma_+(\cos\theta^*>0) + \hat\sigma_-(\cos\theta^*>0)} \;,
\end{equation}
and the backward top polarization $P_t^B$ is similarly defined 
with $\cos\theta^*<0$.  
The forward-backward asymmetry of the top quark is given by
\begin{equation}
A_{\rm FB}^t = {\hat\sigma(\cos\theta^*>0) - \hat\sigma(\cos\theta^*<0) 
\over \hat\sigma(\cos\theta^*>0) + \hat\sigma(\cos\theta^*<0)}\;,
\end{equation}
which can be defined for either a fixed top quark helicity or 
with the helicities summed.

In the case of the $B$-violating interaction, the helicity 
cross sections are obtained from those for the $L$-violating case 
with the replacement $\cos\theta^* \to -\cos\theta^*$ and $\lambda_{t,\bar t} 
\to -\lambda_{t,\bar t}$.  We have 
\begin{mathletters}
\begin{eqnarray}
{d\hat\sigma^{\rm new}_{+-}\over d\cos\theta^*} 
&=& -{\alpha_s |\lambda''_{31k}|^2 \beta\over 72 \hat s}\,
{(1+\beta) (1+\cos\theta^*)^2 
\over 1+2(m_{\tilde d^k}^2-m_t^2)/\hat s + \cos\theta^*}\;,\\[3\jot]
{d\hat\sigma^{\rm new}_{++}\over d\cos\theta^*} =
{d\hat\sigma^{\rm new}_{--}\over d\cos\theta^*} 
&=& -{\alpha_s |\lambda''_{31k}|^2 \beta m_t^2\over 18 \hat s^2}\,
{\sin^2\theta^*
\over 1+2(m_{\tilde d^k}^2-m_t^2)/\hat s + \cos\theta^*}\;,\\[3\jot]
{d\hat\sigma^{\rm new}_{-+}\over d\cos\theta^*} 
&=& -{\alpha_s |\lambda''_{31k}|^2 \beta\over 72 \hat s}\,
{(1-\beta) (1-\cos\theta^*)^2 
\over 1+2(m_{\tilde d^k}^2-m_t^2)/\hat s + \cos\theta^*}\;.
\end{eqnarray}
\end{mathletters}
The expression for $P_t$ changes sign in addition to the replacement 
of the coupling and mass.  The forward and backward hemispheres should be 
also exchanged.

The contribution of the electroweak process 
$q\bar{q} \rightarrow Z^* \rightarrow t\bar{t}$ 
is known to be small~\cite{pari} and will be neglected in the following 
analysis.  It is important to realize here that there is 
{\it no interference\/} with the QCD amplitude 
because of the orthogonal color structure.  
Thus the effect of the $R$-violating interaction can be sizable 
even for a comparable coupling strength.

We calculate the cross section $\sigma$ for $p\bar{p} \to t\bar{t} +X$ 
by convoluting the parton cross section $\hat{\sigma}$ discussed above 
with the parton distribution functions.  
We use the CTEQ3L parton distribution functions~\cite{cteq} with both 
the renormalization and factorization scales chosen to be $\mu=\hat s^{1/2}$.  
The top quark mass of 175 GeV is used.  
We assume that the QCD correction factors (the $K$ factor) 
to $\sigma^{\rm new}$ and $ \sigma^{0}$ take the same value of 1.7. 

For the purpose of estimating the statistical sensitivity, 
we assume that only the leptonic modes 
$t \to W^+ b \to \ell^+\nu_\ell b$ ($\ell=e$ or $\mu$)
are used to identify the top quark.  
This is because charge measurement is needed for top/antitop 
identification and the reconstruction of the final state decay products 
is required for separating the top polarization states from decay 
angular distribution~\cite{pola}.  
The statistical error for the top polarization is given by
\begin{equation}\label{error}
\delta P_t=\frac{\bigl[2(N_+^2 + N_-^2)\bigr]^{1/2}}{(N_+ + N_-)^{3/2}}
               \simeq \frac{1}{ \sqrt{N_+ + N_-} }\;,
\end{equation}
where $N_+$ and $N_-$ are the numbers of right-handed and left-handed 
top events.

The polarization of $\bar t$ can also be used to increase the statistical 
sensitivity of the data.  The total polarizations $P_{\bar t}$ and $P_t$ 
have the same magnitude but opposite in sign within our approximation,
and the hemisphere $\bar t$ polarization can be calculated from the 
expressions given above.  To be conservative, we do not use the $\bar t$ 
polarization in the following
analyses.  

We assume for simplicity that only one of the couplings dominates.  
For the $L$-violating couplings,  $\lambda'_{131}$ and 
$\lambda'_{231}$ are already strongly constrained by atomic parity 
violation and $\nu_{\mu}$ deep-inelastic scattering~\cite{apv}, 
respectively.   
We may thus interpret our result as that on $\lambda'_{331}$, 
in which case the scalar-tau is exchanged in the process.  
In the case of $B$-violating interactions, none of
the relevant couplings have been well constrained by other processes.  
 
In Fig.~2, we show the polarizations $P_t$, $P_t^F$, and $P_t^B$ 
normalized to $|\lambda'_{331}|^2$ 
versus the scalar-tau mass.    
The figures can be also read as  $-P_t/|\lambda''_{31i}|^2$
versus the squark mass with `forward' and `backward' interchanged.   
If the coupling is of the order of the 
weak SU(2) gauge coupling, a polarization of several \%\ is expected, 
without contradicting with existing limits.  
The total polarization is smaller than hemisphere polarizations because 
of the cancellation between the two hemispheres.  
For the coupling $\lambda'_{331}$ ($\lambda''_{31i}$),
the backward (forward)
polarization has the largest magnitude of the three.  In spite of 
the smaller number of events, the statistical sensitivity of the 
backward (forward) polarization is higher than that of the total polarization, 
as may be seen from Fig.~3.  
Because of the small electroweak contribution (much less than one per cent), 
the top polarization can
be a useful probe of the $R$-violating couplings.  

The forward-backward asymmetries for the top quark 
(for individual helicities and their sum)
are shown in Fig.~4.  The size of the effect is 
similar to the polarizations.  The total asymmetry 
is smaller than the asymmetries of each helicity top events.  

The $2\sigma$ limit for $\lambda'_{331}$ 
($\lambda''_{31i}$), which is defined% 
\footnote{Here we neglect the possible systematic error and 
only consider the statistical error.}
by Eq.~(\ref{error}) with $P_t^B/\delta P_t^B = 2$ ($P_t^F/\delta P_t^F = 2$), 
versus scalar-tau (squark) mass is shown in Fig.~5 for the integrated 
luminosities of 2 fb$^{-1}$ (Run 2) and 30 fb$^{-1}$ (Run 3).  
The current $2\sigma$ limits from $Z$ decay at LEP-I~\cite{Zdecay}  
are also shown for comparison.  

Note that the limits on $\lambda'_{331}$ 
obtained from the decay $Z\rightarrow\tau^+\tau^-$ 
depend on squark mass since the coupling  
$\lambda'_{331}$ contributes to this decay via the squark-quark
loops. But our limits depend on the scalar-tau mass. 
Therefore, $P_t$ and $R_{\tau}$ can provide complementary information 
for the coupling $\lambda'_{331}$, and the upgraded 2 TeV Tevatron collider
with $L=30$ fb$^{-1}$ or even lower luminosity can provide meaningful
information on this kind of new physics.  As the polarization grows with 
energy, one can probe further parameter space in the couplings and 
masses at higher energies.

\section{Summary}

We have studied the possibility to probe the $R$-parity violating 
interactions of the top quark at the Tevatron collider.  
Obviously, $R$-violating decays of squarks provide a way to limit 
these interactions~\cite{Allanach}.  We have shown that even when the 
squark is rather heavy and the production cross section is low, 
the $R$-violating interactions can affect the top quark pair production 
cross section via slepton (or squark) exchange.  As these couplings are 
chiral, their contributions can lead to sizable 
polarization and forward-backward 
asymmetry for the top quark, that exceed the level expected from 
the standard model electroweak interactions.  
Hence, we expect that
the forthcoming new runs at the Tevatron collider with main injector 
can improve the constraints on those couplings.

\section*{Acknowledgments}

JMY acknowledges JSPS for financial support.
This work is supported in part by the 
Grant-in-Aid for Scientific Research (No.~10640243) and Grant-in-Aid 
for JSPS Fellows (No.~97317) from the Japan Ministry of Education, 
Science, Sports, and Culture.  BLY acknowledges the support by a 
NATO collaborative grant.
BLY would also like to thank Professor Zhongyuan Zhu and Professor
Ting-Kuo Lee, respectively at the Institute of Theoretical Physics, 
Chinese Academy of Sciences, Beijing and the National Center of 
Theoretical Sciences, Hsin Chu, for their kind hospitalities and
support.

\section*{appendix}
\appendix

Here we list the amplitudes for $d\bar d\to t\bar t$ used in 
the paper.  The Feynman amplitudes are as follows:

\begin{mathletters}
$s$-channel gluon exchange
\begin{equation}
{\cal M}_s = {g_s^2\over s}\;\bar u(p)\gamma^\mu v(\bar p)\;
\bar v(\bar k)\gamma_\mu u(k)\;
\bigl(T^a\bigr)_{\gamma\delta} \bigl(T^a\bigr)_{\beta\alpha} 
\end{equation}
\indent $t$-channel slepton exchange
\begin{equation}
{\cal M}_t = {|\lambda'|^2\over t-m_{\tilde\ell}^2}\;
\bar u(p) {\textstyle{1+\gamma_5\over 2}} u(k)\;
\bar v(\bar k) {\textstyle{1-\gamma_5\over 2}} v(\bar p)\;
\delta_{\gamma\alpha} \delta_{\delta\beta} 
\end{equation}
\indent $u$-channel squark exchange
\begin{equation}
{\cal M}_u = {|\lambda''|^2\over u-m_{\tilde q}^2}\;
\bar u(p) {\textstyle{1-\gamma_5\over 2}} C \bar v^T(\bar k)\;
u^T(k) C^\dagger {\textstyle{1+\gamma_5\over 2}} v(\bar p)\;
\epsilon_{\delta\alpha\epsilon} \epsilon_{\gamma\beta\epsilon} 
\end{equation}
\end{mathletters}
In these formulas, $\lambda'$ and $\lambda''$ are to be interpreted 
as $ \lambda'_{i31}$ and $\lambda''_{31k}$.

After Fierz rearrangement, crossed channel amplitudes can be 
written in the $s$-channel form
\begin{mathletters}
\begin{eqnarray}
{\cal M}_t &=& {|\lambda'|^2\over t-m_{\tilde\ell}^2}\;
\bar u(p) \gamma^\mu {\textstyle{1-\gamma_5\over 2}} v(\bar p)\;
\bar v(\bar k) \gamma_\mu {\textstyle{1+\gamma_5\over 2}} u(k)\;
\delta_{\gamma\alpha} \delta_{\delta\beta} \\
{\cal M}_u &=& -{|\lambda''|^2\over u-m_{\tilde q}^2}\;
\bar u(p) \gamma^\mu {\textstyle{1+\gamma_5\over 2}} v(\bar p)\;
\bar v(\bar k) \gamma_\mu {\textstyle{1+\gamma_5\over 2}} u(k)\;
\delta_{\gamma\alpha} \delta_{\delta\beta} 
\end{eqnarray}
\end{mathletters}

The helicity amplitudes for the process may be calculated from 
these invariant amplitudes.  The amplitude for the helicity 
combination 
\begin{equation}
d(\lambda_d) + \bar d(\bar\lambda_d) \to 
t(\lambda_t) + \bar t(\bar\lambda_t)
\end{equation}
has the structure
\begin{equation}
{\cal M} = \widetilde{\cal M}(E_{\rm cm},\cos\theta)\; 
d^1_{\lambda_i\lambda_f}(\theta)\; e^{i(\lambda_i-\lambda_f)\varphi} 
\end{equation}
where $\lambda_i=\lambda_d - \bar\lambda_d$, 
$\lambda_f=\lambda_t - \bar\lambda_t$, $(\theta,\varphi)$ the 
scattering angle (the direction of the final $t$ with respect to 
the initial $d$), and $d^1_{\lambda\lambda'}$ is the Wigner d function 
(rotation matrix) for the spin-1 representation.  

Since chirality is conserved, the process occurs for only two of 
the initial states with $\lambda_i=\pm1$.  
The standard model contribution is   
\begin{equation}
\widetilde{\cal M}_s = \cases{
-g_s^2 C_s & $\lambda_f=\pm1$\cr
\displaystyle -{1\over\sqrt2\gamma}\,g_s^2 C_s & $\lambda_f=0$\cr}
\end{equation}
The slepton exchange contribution exists only for $\lambda_i=+1$ 
because of the chirality structure of the interaction (only $d_L$ 
interacts):  
\begin{equation}
\widetilde{\cal M}_t = \cases{\displaystyle
-|\lambda'|^2\;{s\over 2(t-m_{\tilde e}^2)}\;(1\mp\beta) C_t 
&$\lambda_f=\pm1$\cr
\displaystyle
-|\lambda'|^2\;{s\over 2(t-m_{\tilde e}^2)} 
\times{1\over\sqrt2\gamma} C_t
&$\lambda_f=0$\cr}
\end{equation}
The squark exchange contribution is also for $\lambda_i=+1$ only:
\begin{equation}
\widetilde{\cal M}_u = \cases{\displaystyle
-|\lambda''|^2\;{s\over 2(u-m_{\tilde d}^2)}\;(1\pm\beta) C_u
&$\lambda_f=\pm1$\cr 
\displaystyle
-|\lambda''|^2\;{s\over 2(u-m_{\tilde d}^2)} 
\times{1\over\sqrt2\gamma} C_u
&$\lambda_f=0$\cr}
\end{equation}
where $\beta=(1-4m_t^2/s)^{1/2}$ is the top c.m.\ velocity, 
$\gamma=(1-\beta^2)^{-1/2} = \sqrt s/2m_t$, 
and the color factors are given by
\begin{eqnarray}
C_s &=& \delta_{\alpha\gamma}\delta_{\beta\delta}
- {1\over3} \delta_{\alpha\beta}\delta_{\gamma\delta}\\
C_t &=& \delta_{\alpha\gamma}\delta_{\beta\delta}\\
C_u &=& \delta_{\alpha\gamma}\delta_{\beta\delta}
-  \delta_{\alpha\beta}\delta_{\gamma\delta}
\end{eqnarray}

%%%%%%%%%%%%%%%%%%%%%%%%%
\begin{figure}
\label{fig1}
\vfil
\begin{center}
%Fig.~1 
\psfig{figure=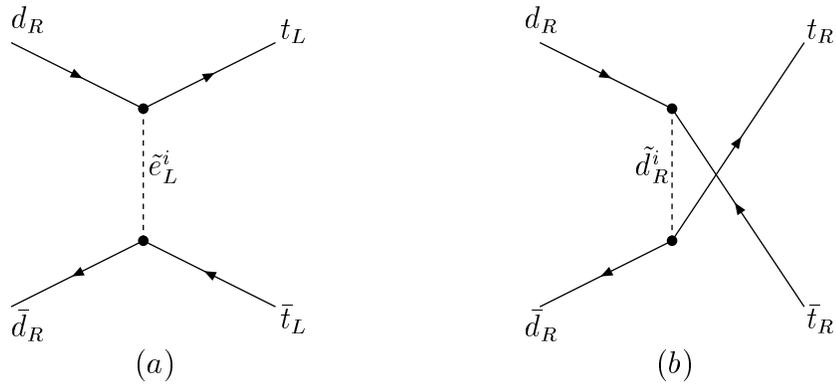,width=400pt,angle=0}
\end{center}
\caption{
% Fig.1 
Feynman diagrams for (a) the $L$-violating process 
$d\bar d\rightarrow t_L \bar t_L$ induced by  $\lambda'_{i31}$,  
and (b) the $B$-violating process $d\bar d\rightarrow t_R \bar t_R$
induced by $\lambda''_{31i}$.}
\vfil\vfil
\end{figure}
%%%%%%%%%%%%%%%%%%%%%%%%%%
\begin{figure}
\label{fig2}
\hrule height 0pt
\vfil
\begin{center}
%Fig.~2 
\psfig{figure=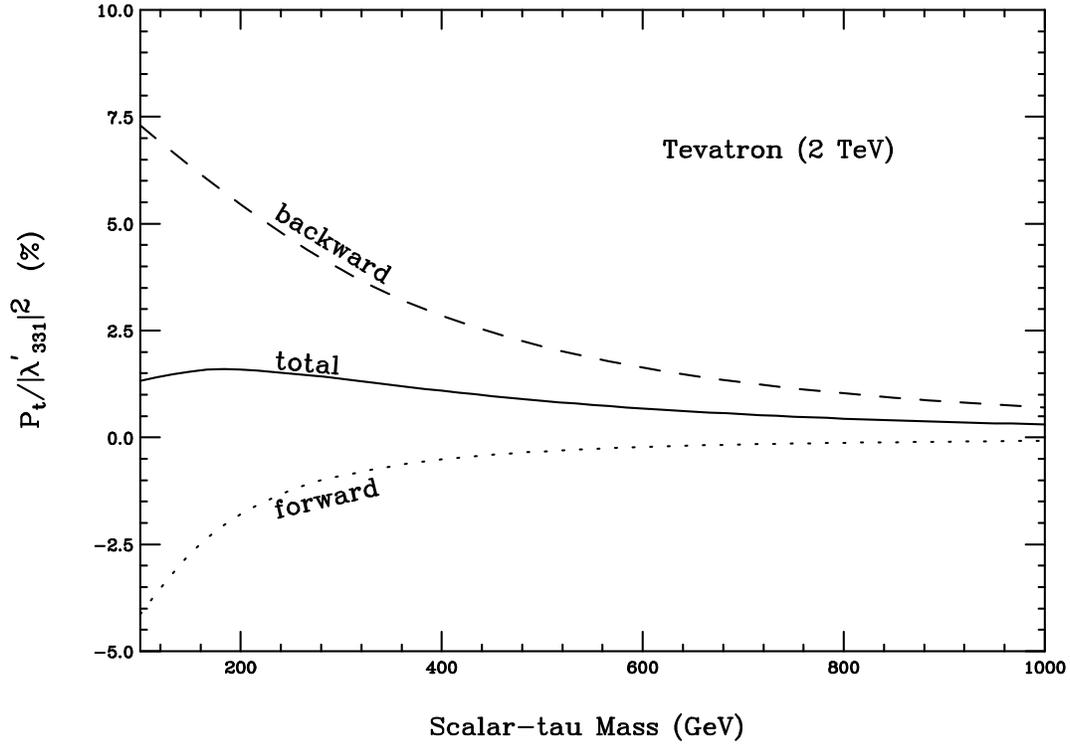,width=400pt,angle=90}
\end{center}
\hspace*{1cm}
\caption{
% Fig.2 
Top quark polarizations at $\sqrt s=2$ TeV 
versus the exchanged scalar-tau mass.  
Total: $P_t$; forward: $P_t^F$: backward: $P_t^B$.
The figure can be also read as  $-P_t/|\lambda''_{31i}|^2$
versus the  exchanged squark mass with `forward' and `backward'
interchanged. 
}
\vfil\vfil
\end{figure}
%%%%%%%%%%%%%%%%%%%%%%%%%%
%%%%%%%%%%%%%%%%%%%%%%%%%
\begin{figure}
\label{fig3}
\hrule height 0pt
\vfil
\begin{center}
%Fig.~3
\psfig{figure=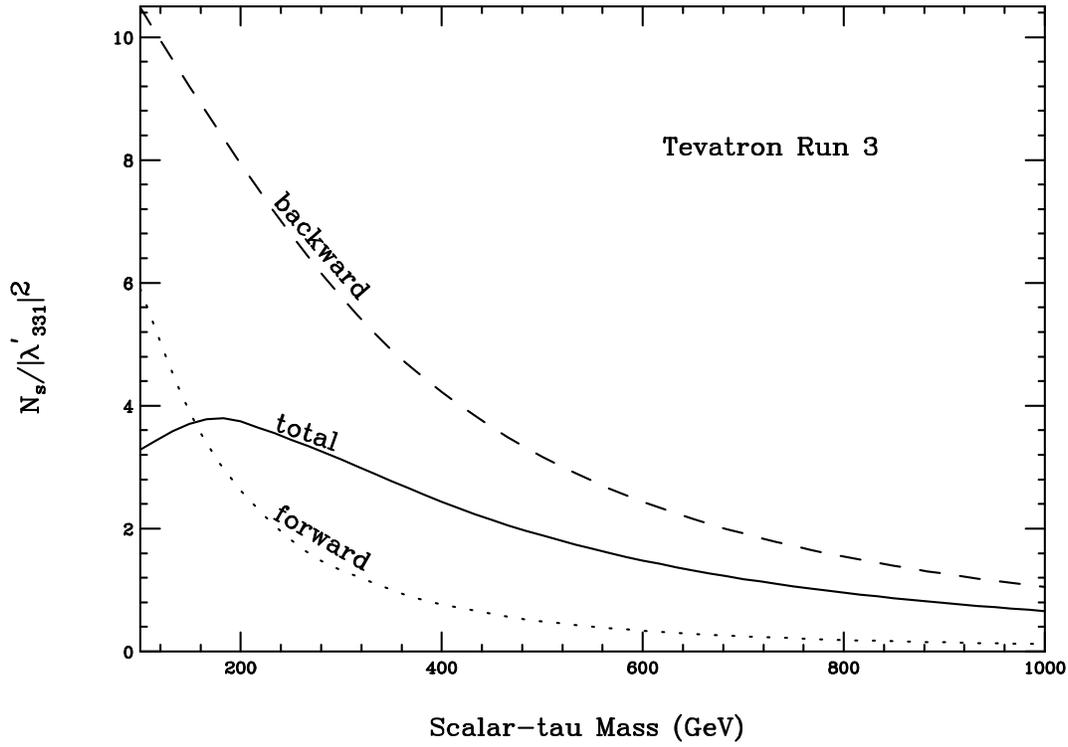,width=400pt,angle=90}
\end{center}
\hspace*{1cm}
\caption{
%Fig.3
Statistical significance at $\sqrt s=2$ TeV with $L=30$ fb$^{-1}$.
The figure can be also read as  $N_s/|\lambda''_{31i}|^2$
versus the  exchanged squark mass with `forward' and `backward'
interchanged. 
}  
\vfil\vfil
\end{figure}
%%%%%%%%%%%%%%%%%%%%%%%%%%
%%%%%%%%%%%%%%%%%%%%%%%%%
\begin{figure}
\label{fig4}
\hrule height 0pt
\vfil
\begin{center}
%Fig.~4
\psfig{figure=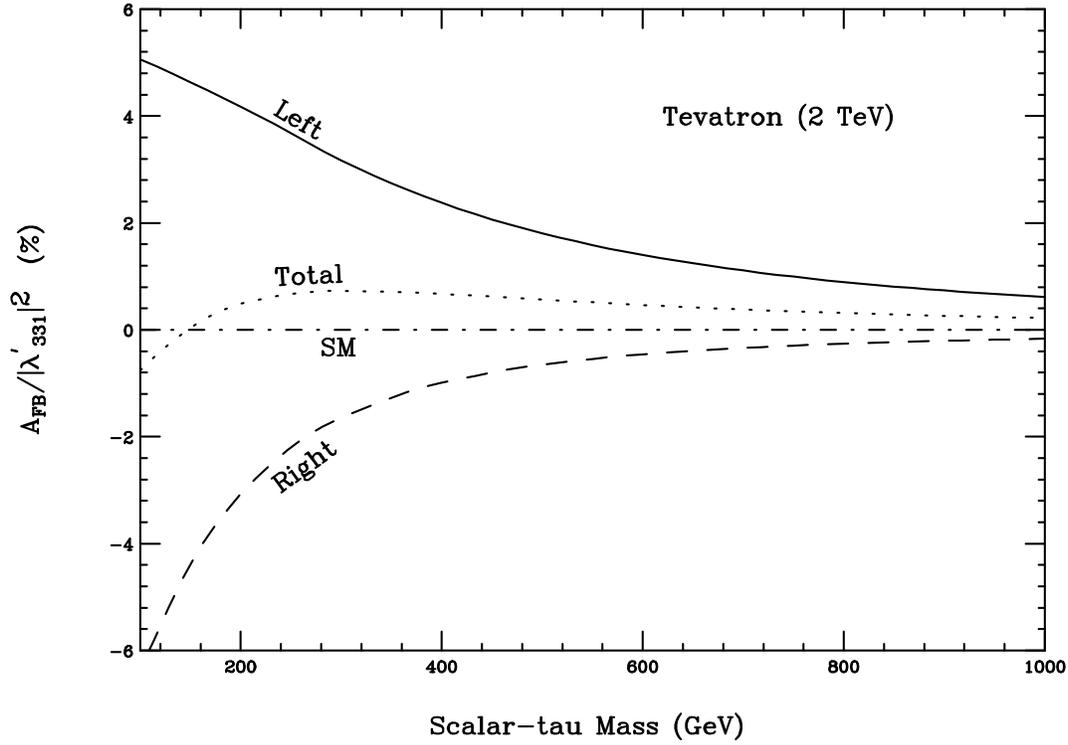,width=400pt,angle=90}
\end{center}
\hspace*{1cm}
\caption{
%Fig.4
Top quark forward-backward asymmetry at $\sqrt s=2$ TeV 
versus the exchanged sfermion mass for all events (total); 
left-handed (left) and right-handed (right) top events.
The figure can be also read as  $-A_{FB}/|\lambda''_{31i}|^2$
versus the  exchanged squark mass with `Left' and `Right'
interchanged. }  
\vfil\vfil
\end{figure}
%%%%%%%%%%%%%%%%%%%%%%%%%%
%%%%%%%%%%%%%%%%%%%%%%%%%
\begin{figure}
\label{fig5}
\hrule height 0pt
\vfil
\begin{center}
%Fig.~5
\psfig{figure=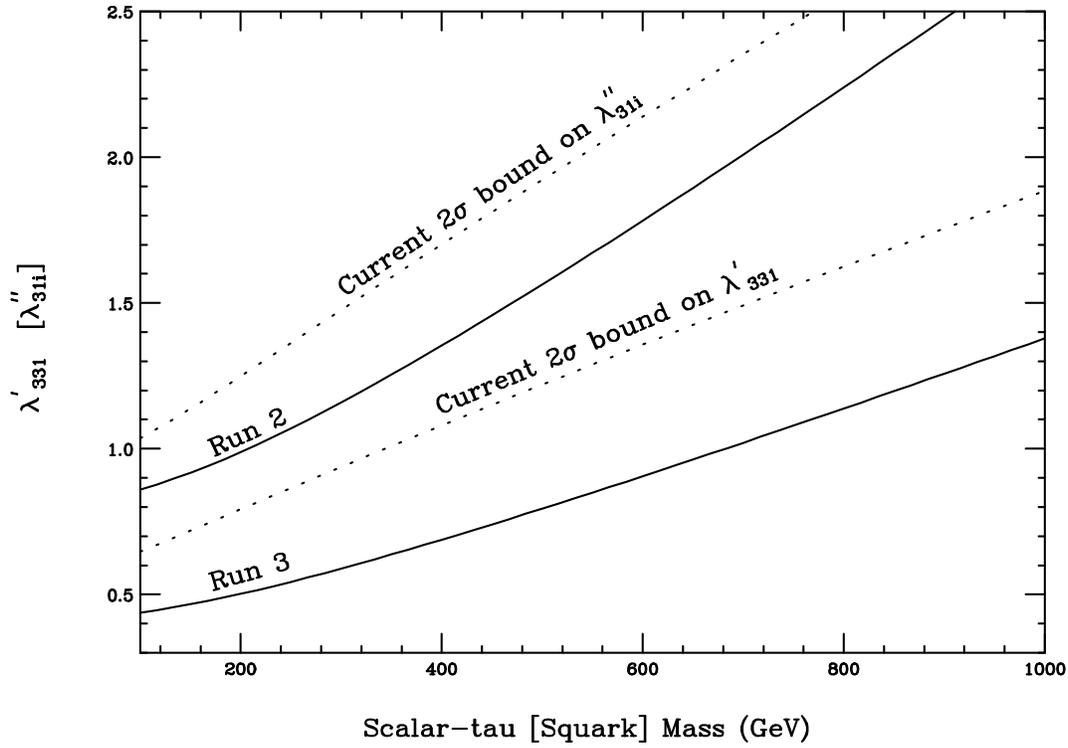,width=400pt,angle=90}
\end{center}
\hspace*{1cm}
\caption{
%Fig.5
The expected $2\sigma$ limit for $\lambda'_{331}$ 
($\lambda''_{31i}$) versus scalar-tau (squark) mass.
The two solid curves are at $\sqrt s=2$ TeV with 
$L=2$ fb$^{-1}$ (Run 2) and 30 fb$^{-1}$ (Run 3), respectively. 
The current limits from $Z$ decay are also shown as a function 
of the squark mass.}
\vfil\vfil
\end{figure}
%%%%%%%%%%%%%%%%%%%%%%%%%%

\end{document}